\documentclass[aps,prl,twocolumn,groupedaddress,longbibliography]{revtex4-2}

\usepackage{booktabs}
\usepackage{amsmath}
\usepackage{graphicx}
\usepackage{verbatim}
\usepackage{subfigure}
\usepackage[english]{babel}
\usepackage{verbatim}
\usepackage{multirow}
\usepackage{xcolor}
\usepackage[utf8]{inputenc}
\usepackage{braket}
\usepackage{bm}
\usepackage{float} 
\usepackage{tabularx}

\begin{document}

\title{Precision spectroscopy of the fine and hyperfine structures of high molecular Rydberg-Stark states: Metrology of molecular hydrogen ions}

\author{I. Doran\textsuperscript{1}}

\author{L. Jeckel\textsuperscript{1}}

\author{M. Beyer\textsuperscript{2}}
\author{Ch. Jungen\textsuperscript{3}}

\author{F. Merkt\textsuperscript{1,4,5}} 
\email{frederic.merkt@phys.chem.ethz.ch (F. Merkt)} 

\affiliation{%
  \textsuperscript{1}Institute of Molecular Physical Science, ETH Z\"{u}rich, 8093 Z\"{u}rich, Switzerland%
}

\affiliation{%
  \textsuperscript{2}Department of Physics and Astronomy, LaserLaB, Vrije Universiteit Amsterdam, de Boelelaan 1081, 1081 HV Amsterdam, The Netherlands%
}

\affiliation{%
  \textsuperscript{3}Universit\'{e} Paris-Saclay, Centre National de la Recherche Scientifique, Laboratoire Aim\'{e} Cotton, 91400 Orsay, France
}

\affiliation{%
  \textsuperscript{4}Department of Physics, ETH Z\"urich, Z\"urich, Switzerland}
  
\affiliation{%
  \textsuperscript{5}Quantum Center, ETH Z\"urich, Z\"urich, Switzerland}

\date{\today}

\begin{abstract}
The Stark effect in autoionizing high-$n$ Rydberg states decouples the Rydberg electron from the ion core through $\ell$ mixing with core-nonpenetrating high-$\ell$ states. The Rydberg states become long-lived, which is ideal for precision spectroscopy, and their structures reflect the fine and hyperfine structures of the ion-core levels. We report on precision measurements, in weak electric fields, of the fine and hyperfine structures of two distinct categories of high autoionizing molecular Rydberg-Stark states 
differing by the nature of the ion-core angular momentum: Rydberg states of para-H$_2$ (total nuclear spin $I=0$) with a rotationally excited ($N^+=2$) H$_2^+$ ion core and Rydberg states of ortho-D$_2$ ($I=2$) with a rotationless ($N^+=0$) ion core. The spectra reveal striking differences which are interpreted as arising from the dominance of anisotropic charge-quadrupole interactions between the rotating quadrupolar ion core and the Rydberg electron in para-H$_2$ and the absence of such interactions in rotationless ortho-D$_2$ Rydberg states. In ortho-D$_2$, the dominant interaction, the magnetic Fermi-contact hyperfine interaction in the ion core, does not significantly affect the motion of the Rydberg electron.
By analyzing these spectra based on a treatment combining multichannel quantum-defect theory and matrix diagonalization, we derive new experimental values of the hyperfine coupling constant $b_F$ = 139.84(5)~MHz of D$_2^+ (v^+=1, N^+=0)$, the spin-rotation coupling constant $c_e$ = 39.62(11)~MHz of H$_2^+ (v^+=1, N^+=2)$ and the fundamental vibrational interval of ortho-D$_2^+$ (47\,279\,980.8(1.9)~MHz). 
The approach followed here in the study of molecular Rydberg-Stark states is general and broadly applicable to measurements of the fine and hyperfine structures of molecular cations.

\end{abstract}

\maketitle

\begin{figure*}
	\centering
	{\includegraphics[trim=0cm 0cm 0cm 0cm, clip=true, width=1.0\linewidth]{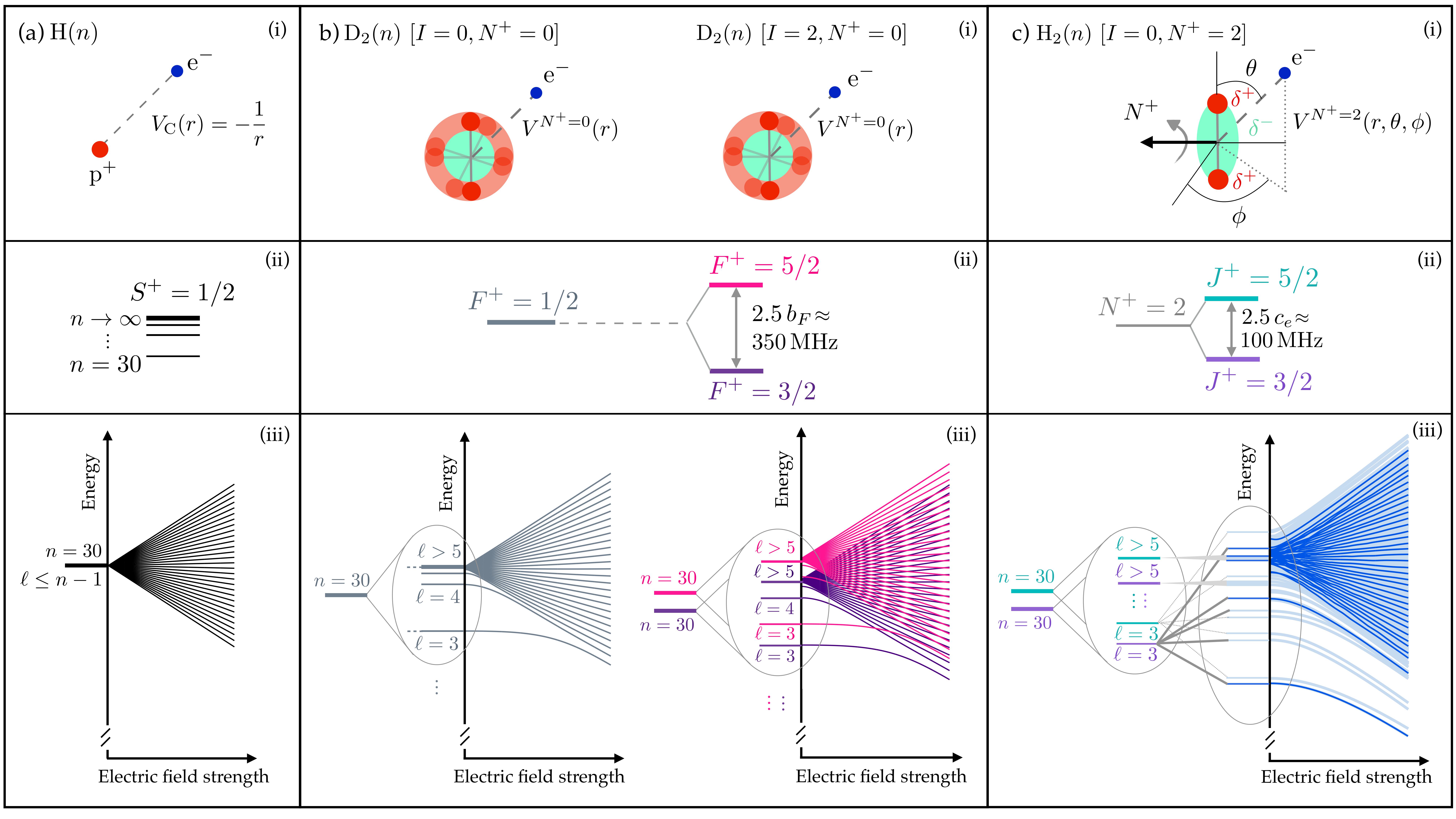}}
	\caption{ Schematic illustration of the salient features of the Stark effect in Rydberg states of (a) the H atom with a point-like H$^+$ ion core and a pure Coulomb potential $V_{\rm C}(r)$, (b) ortho-D$_2$ with a $N^+=0$ ion core and a central potential ($V^{N^+=0}(r)$) deviating at short range from a Coulomb potential. Left: $I=0$; right: $I=2$ (with core hyperfine structure), and (c) para-H$_2$ with $N^+=2$ with spin-rotation splitting and a quadrupolar ion core causing an anisotropic long-range charge-quadrupole interaction $V^{N^+=2}(r,\theta,\phi)$. (i) Rydberg-electron--ion-core interactions. The ion-core quadrupole is indicated schematically by the partial charges and their color code in (b) and (c). (ii) Ion-core level structure. (iii) Rydberg-level structure at $n=30$ and its field dependence. See text for details.}
	\vspace{-0.5cm}
	\label{fig1}
\end{figure*}

We report on precision measurements of the spectra of highly excited electronic states (Rydberg states) of the two-electron molecules H$_2$ and D$_2$ in the presence of external electric fields. In these states, the molecules can be regarded as consisting of an H$_2^+$ or D$_2^+$ ion core and a weakly bound electron moving in distant orbits corresponding to large values of the principal quantum number $n$, as illustrated schematically in Fig.~\ref{fig1}. Even weak electric fields induce a large Stark effect in these states and strongly modify their structure and dynamics \cite{gallagher94a,stebbings83a}.  The measurements aim at characterizing all interactions in these highly excited states, including those involving the magnetic and electric moments of the nuclei (leading to the hyperfine structure), the ion-core molecular rotation and the spin and orbital motion of the excited electron (leading to the fine structure).

The precise characterization of the structure and dynamics of simple, few-body atomic and molecular systems is of intrinsic fundamental interest, and also increasingly used to improve the values of fundamental constants, test fundamental physical theories and search for physics beyond the standard model of particle physics \cite{germann21a,delaunay23a,korobov25a,alighanbari25a,schiller24a,puchalski19a,hoelsch19a,patkos21a,clausen25b,mohr25a}. Current efforts are also invested towards exploiting high atomic and molecular Rydberg-Stark states in electric-field quantum sensing \cite{facon16a,zhang24a,liang26a} and quantum-information processing \cite{nipper12a,weber17a,stecker20a,jiao22a,mehaignerie25a}. Rydberg states exhibiting hyperfine structures are attractive systems in this context because 
nuclear spins can be exploited to enhance the efficiency of quantum gates and quantum-information-processing protocols \cite{peper25a}.

The Stark effect in high atomic and molecular Rydberg states is particularly strong in energy regions near zero-quantum-defect positions, where $\ell \geq 4$ states are almost degenerate ($\ell$ is the Rydberg-electron orbital-angular-momentum quantum number). 
The Stark shifts are linear even at low field strengths \cite{bethe57a,stebbings83a,gallagher94a} and the level structure fans out with increasing field strength $\mathcal{F}$ in linear manifolds of Stark states see Fig.~\ref{fig1}. The Stark states are labeled with an index $k$ ranging from $-(n-|m|-1)$ to $(n-|m|-1)$ in steps of 2. The linear Stark shifts are $3nk\mathcal{F}/2$ (in atomic units), giving rise to characteristic patterns of equidistant states \cite{bethe57a,stebbings83a,gallagher94a}.
The electric field decouples the Rydberg electron from the ion core by inducing $\ell$ mixing with long-lived core-nonpenetrating high-$\ell$ Rydberg states ($\ell\geq 4$). Three advantages result for precision spectroscopy: (i) the Rydberg states become long-lived and the autoionization resonances sharp, (ii) the ion-core level structure can be deduced from the structure of the Rydberg-Stark states \cite{doran24a}, and (iii) $\ell$ mixing provides access to multiple Rydberg series and to a global understanding of their structure and dynamics.

In Rydberg states of atoms with closed-shell ion cores, the Stark effect can be accurately treated by matrix-diagonalization procedures \cite{zimmerman79a,grimmel15a,sibalic17a,peper19a,scheidegger23a}. These procedures can be extended to Rydberg atoms with open-shell ion cores (see Refs.~\cite{ernst88a,brevet90a,fielding92a,vrakking97a} for early examples) and molecules \cite{bordas87a,bordas92a,fielding91a,vrakking96a,seiler11b,rayment21a,munkes24a}. Alternative treatments of the Stark effect based on multichannel quantum-defect theory have also been developed \cite{harmin81a,harmin84a,sakimoto86a,sakimoto89a,fielding91b,softley97b,gruetter08a,giannakeas16a}. The understanding of the Stark effect in Rydberg states is thus satisfactory, with one important exception: No experimental data have been reported yet on molecular Rydberg-Stark states at a sufficiently high resolution to reveal their hyperfine structures. This gap is filled in the present letter.

\begin{figure*}
	\centering
	{\includegraphics[trim=8cm 12cm 4cm 2cm, clip=true, width=\linewidth]{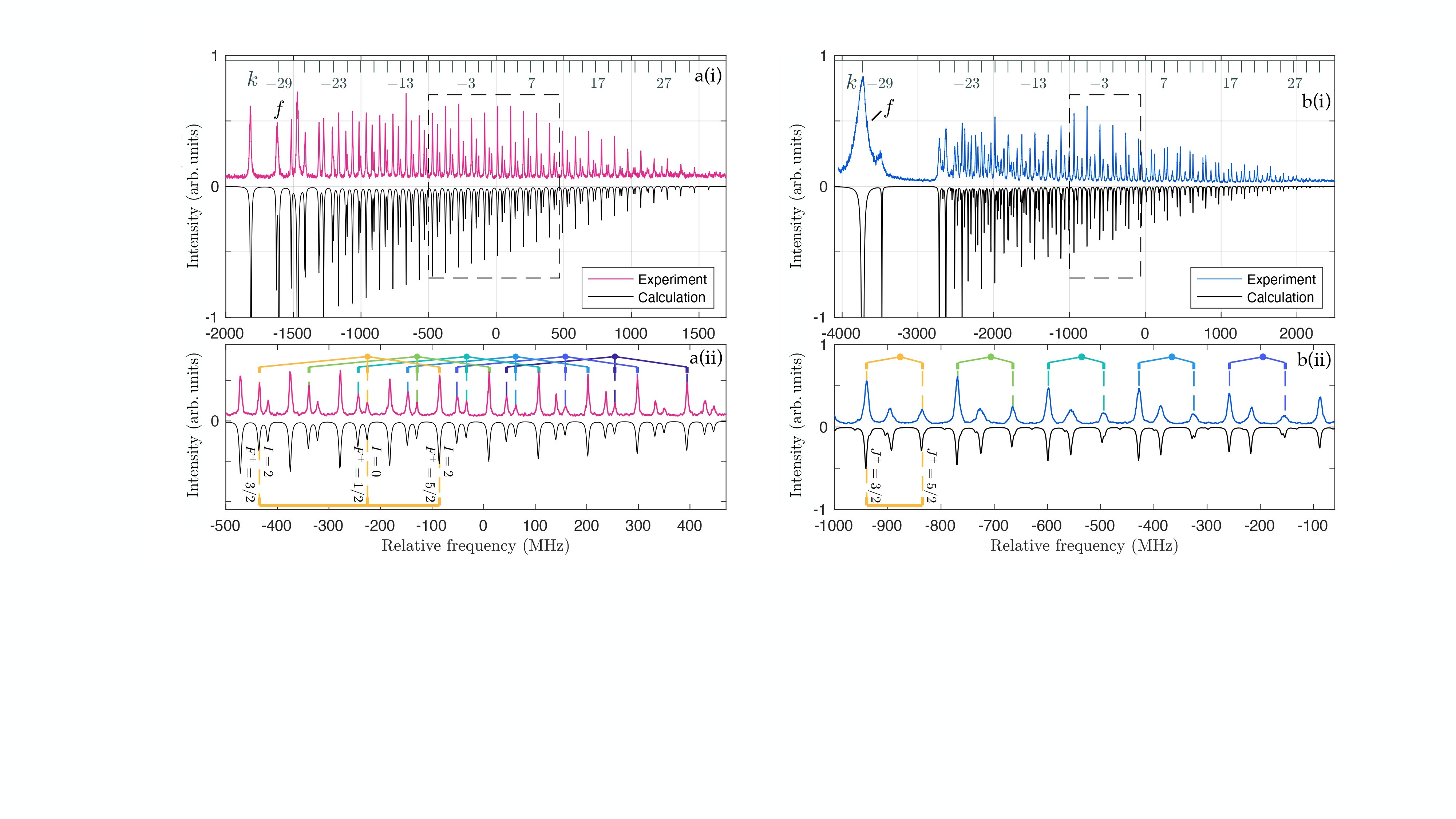}}
	\caption{Measured spectra of the $n=34$ [a: D$_2^+$ X$^+$(1,0) and b: H$_2^+$ X$^+$(1,2)] Rydberg-Stark manifolds at electric fields of 69.79 and 124.35 V/m, respectively, as determined from fits, and calculated spectra (inverted traces) assuming line widths of 5 MHz for all lines. Selected spectral regions [delimited by dashed lines in panels (i)] are enlarged in panels (ii). See text.}
	\vspace{-0.5cm}
	\label{fig:spectra_calc_exp}
\end{figure*}
When the ion core exhibits coupled electronic ($\Lambda^+,\Omega^+$), rotational ($N^+$), electron-spin ($S^+$) and nuclear-spin ($I$) degrees of freedom, the rigorous treatment of the Stark effect necessitates the systematic use of angular-momentum tensor algebra and elaborate representations of complex angular-momentum coupling hierarchies \cite{jungen98a}. These hierarchies evolve rapidly with $n$ and $\ell$ and are strongly modified through $\ell$ mixing by external electric fields. 
We have recently developed a new approach \cite{iodo_tbp_first} combining multichannel quantum-defect theory \cite{jungen96a,jungen11a} and matrix diagonalization \cite{zimmerman79a} to accurately calculate the fine and hyperfine structures of molecular Rydberg-Stark states with open-shell ion cores. To test this approach, we present here studies of two categories of Rydberg-Stark states of H$_2$ and D$_2$ with ion cores in the first vibrationally excited state ($v^+=1$) of the ground X$^+$ $^2\Sigma_g^+$ electronic state. These states
decay by autoionization to the available continua. However, the rapid  autoionization of low-$\ell$ Rydberg states is suppressed in Stark states by field-induced $\ell$ mixing, which is crucial for the observation of their fine and hyperfine stuctures. 
A schematic representation and a comparison with the more familiar situation encountered in the H atom is presented in Fig.~\ref{fig1}. 

The first category (Fig.~\ref{fig1}b) concerns Rydberg states of ortho-D$_2$ ($I=0,2$) with a rotationless $(N^+=0)$ D$_2^+$ ion core. The D$_2^+$ ion core splits into one $I=0,S^+=F^+=1/2$ component and two $I=2$ hyperfine components with $F^+=3/2$ and $F^+=5/2$, separated by 2.5$b_F$ \cite{danev21a}. The second category (Fig.~\ref{fig1}c) concerns Rydberg states of para-H$_2$ ($I=0$) with a $N^+=2$ H$_2^+$ ion core, which splits into two ($J^+=3/2$ and $J^+=5/2$) fine-structure (spin-rotation) components separated by 2.5$c_e$ \cite{korobov06c} (Fig.~\ref{fig1}c).
The top part of Fig.~\ref{fig1} illustrates the interaction of the Rydberg electron with (a) a proton, (b) a nonrotating ({\it i.e.}, spherically symmetric) quadrupolar and polarizable molecular ion core, and (c) a rotating quadrupolar and polarizable molecular ion core. The middle parts (ii) depict the ion-core level structures, and the bottom parts (iii) 
show the level structures of the Rydberg states with (right) and without (left) field. 

The $I=0$ states in ortho-D$_2$ enable the direct comparison with atomic hydrogen (Fig.~\ref{fig1}a) and the identification, by comparison with the $I=2$ nuclear-spin-symmetry species, of the effects arising from the hyperfine structure. 
In both categories, the ion core comprises a single angular momentum [$N^+=2$ in the case of H$_2^+$ ($I=0$) and $I=2$ in the case of D$_2^+$ ($N^+=0$)] in addition to the core-electron spin.  These systems turned out to be ideal to validate our new computational approach \cite{iodo_tbp_first} and enabled the observation of phenomena not anticipated at the outset of the investigation: (i) strikingly different spectral patterns in H$_2$ and D$_2$, which could be interpreted during the analysis, and (ii) different autoionization dynamics of different hyperfine components of specific Stark states in D$_2$, which remains unexplained. These systems were also ideal to demonstrate the possibility of precisely determining the fine and hyperfine structures of molecular ions from Rydberg-Stark spectra. 

The experimental procedure relies on the zero-quantum-defect method \cite{hoelsch22a, doran24a} and the experimental setup described in Refs. \cite{beyer18a, doran24a}. In brief, Rydberg states converging to the ($v^+=1, N^+=0$) and ($v^+=1, N^+=2$) levels of the X$^+$ $^2\Sigma_g^+$ electronic ground state of D$_2^+$ and H$_2^+$, respectively, are accessed from the ($v=2,N=2$) and ($v=2,N=0$) levels of the GK $^1\Sigma_g^+$ double-minimum electronically excited state of H$_2$ using cw single-mode near-infrared (NIR) laser radiation. These GK levels are prepared by resonant two-photon excitation from the ground state using pulsed vacuum-ultraviolet and visible laser radiation (repetition rate 25 Hz, pulse duration $\approx$ 5 ns) \cite{beyer18a}. To minimize Doppler broadening and shifts, a doubly skimmed supersonic beam emitted from a cryogenic pulsed valve is crossed at near-right angles by the NIR laser beam. Retroreflection of this laser beam gives rise to two Doppler components with opposite Doppler shifts, which are averaged to obtain the first-order-Doppler-free transition frequencies \cite{beyer18a}. To avoid spectral congestion, the final measurements are carried out after blocking the retroreflected NIR laser beam.  
The frequency calibration is performed using a frequency comb (Menlo Systems, FC1500-250-WG) referenced to a 10 MHz Rb GPS standard, see Ref.~\cite{beyer18a}. The photoexcitation region is surrounded by a magnetic shield to reduce stray magnetic fields to below $2\cdot 10^{-7}$~T. To record Stark spectra, weak ($<$150 V/m) homogeneous electric fields $\mathcal{F}$ in the photoexcitation region are generated by applying dc electric potentials $V_\textrm{dc}$ across a stack of cylindrical electrodes \cite{beyer18a}. This stack is also used to apply large pulsed electric potentials $V_\textrm{extr}$, typically $\approx $ 1 $\mu$s after photoexcitation, to extract the ions produced following autoionization of the Rydberg-Stark states towards a microchannel-plate detector. Spectra are recorded by monitoring the autoionization yield as a function of the NIR laser frequency.

To characterize the Stark effect and derive ionization energies and the ion fine and hyperfine structures, we measured spectra between $n \approx$ 30 and 45 and field strengths between $\approx 70$ and 150 V/m for Rydberg states converging to both D$_2^+$ X$^+$(1,0) and H$_2^+$ X$^+$(1,2), see Tables SI and SII of the Supplemental Material \cite{supplemental_doran26b}. 
Fig.~\ref{fig:spectra_calc_exp} shows, as illustrations, Stark spectra of $n=34$ Rydberg-Stark states of D$_2$ (a) and H$_2$ (b).
Both spectra reveal a complex structure, reflecting the fine and hyperfine structures. 
In the D$_2$ spectrum, each Stark state ($k$) appears as a triplet that exactly matches the hyperfine structure of the ion core, see Fig.~\ref{fig1}b(ii), as indicated with the colored assignment bars in Fig.~\ref{fig:spectra_calc_exp}a(ii) for $k$ values between $-3$ and 7. 
In analogy, one would expect the Stark states in the case of H$_2$ to split into two manifolds associated with the $N^+=2, J^+=3/2$ and 5/2 spin-rotation components of the ion core, see Fig.~\ref{fig1}c(ii). This is not the case: The H$_2$ spectrum consists of at least three distinct progressions [see Fig. \ref{fig:spectra_calc_exp}b(ii)]. In selected regions, the fine-structure splittings match the ion-core spin-rotation intervals, as indicated by the colored doublets in Fig.~\ref{fig:spectra_calc_exp}b(ii), whereas in other regions no splittings corresponding to the ion-core spin-rotation interval are observable at all, see, e.g., the region around $-2200$~MHz in Fig.~\ref{fig:spectra_calc_exp}b(i). The H$_2$ spectrum is less regular than the D$_2$ spectrum despite the fact that the H$_2^+$ and D$_2^+$ ion cores each exhibit only one type of angular momentum ($I=2$ for D$_2$ and $N^+=2$ for H$_2$) beside the core-electron spin.

To understand the origin of this difference and assign the spectra, the Stark effect was calculated by diagonalizing the matrix of the total effective Hamiltonian $\hat{H} = \hat{H}_0 + e \mathcal{F} \hat{z}$ \cite{zimmerman79a} consisting of the zero-field Hamiltonian $\hat{H}_0$ and the interaction $e \mathcal{F} \hat{z}$ with the external field. The elements of $\hat{H}_0$ correspond to the zero-field energies depicted in Fig.~\ref{fig1}b(iii) and c(iii) and were determined using multichannel quantum-defect-theory (MQDT) calculations for $\ell \leq 3$ Rydberg states and a long-range interaction model accounting for both electrostatic and hyperfine interactions for $\ell \geq 4$ Rydberg states \cite{iodo_tbp_first}. 
The matrix elements of $e \mathcal{F} \hat{z}$ were evaluated using angular-momentum tensor algebra~\cite{iodo_tbp_first} and lead to the Stark-manifolds depicted schematically in Fig.~\ref{fig1}b(iii) and c(iii). Because the intermediate GK state has dominant singlet d ($S=0,\ell=2$) character, the transition intensities are in good approximation proportional to the singlet f ($S=0,\ell=3$) character of the Rydberg-Stark states. 

Agreement between measured and calculated (inverted traces in Figure~\ref{fig:spectra_calc_exp}) spectra was reached in a fit of the ionization energies [i.e., the D$_2^+$ X$^+$ (1,0) - D$_2$ GK(2,2) and H$_2^+$ X$^+$ (1,2) - H$_2$ GK(2,0) intervals], the electric field strengths [69.79(3) and 124.35(3) V/m for D$_2$ and H$_2$ respectively], as well as the Fermi-contact hyperfine coupling constant $b_F$ in the case of D$_2^+$ X$^+$(1,0) and the spin-rotation coupling constant $c_e$ in the case of H$_2^+$ X$^+$(1,2) (see Table \ref{tab:vals_final} for the final results).  
The calculated line positions reproduce the experimental ones well, with weighted RMS fit values below 5. The intensities are also well reproduced, particularly on the high-frequency side. On the low-frequency side, saturation of the transitions and lifetime broadening of states with large f character lead to reduced experimental peak intensities, see, e.g., the line labeled "\textit{f}" in the H$_2$ spectrum.

\begin{figure}
	\centering
	{\includegraphics[trim=0cm 0cm 0cm 0cm, clip=true, width=\linewidth]{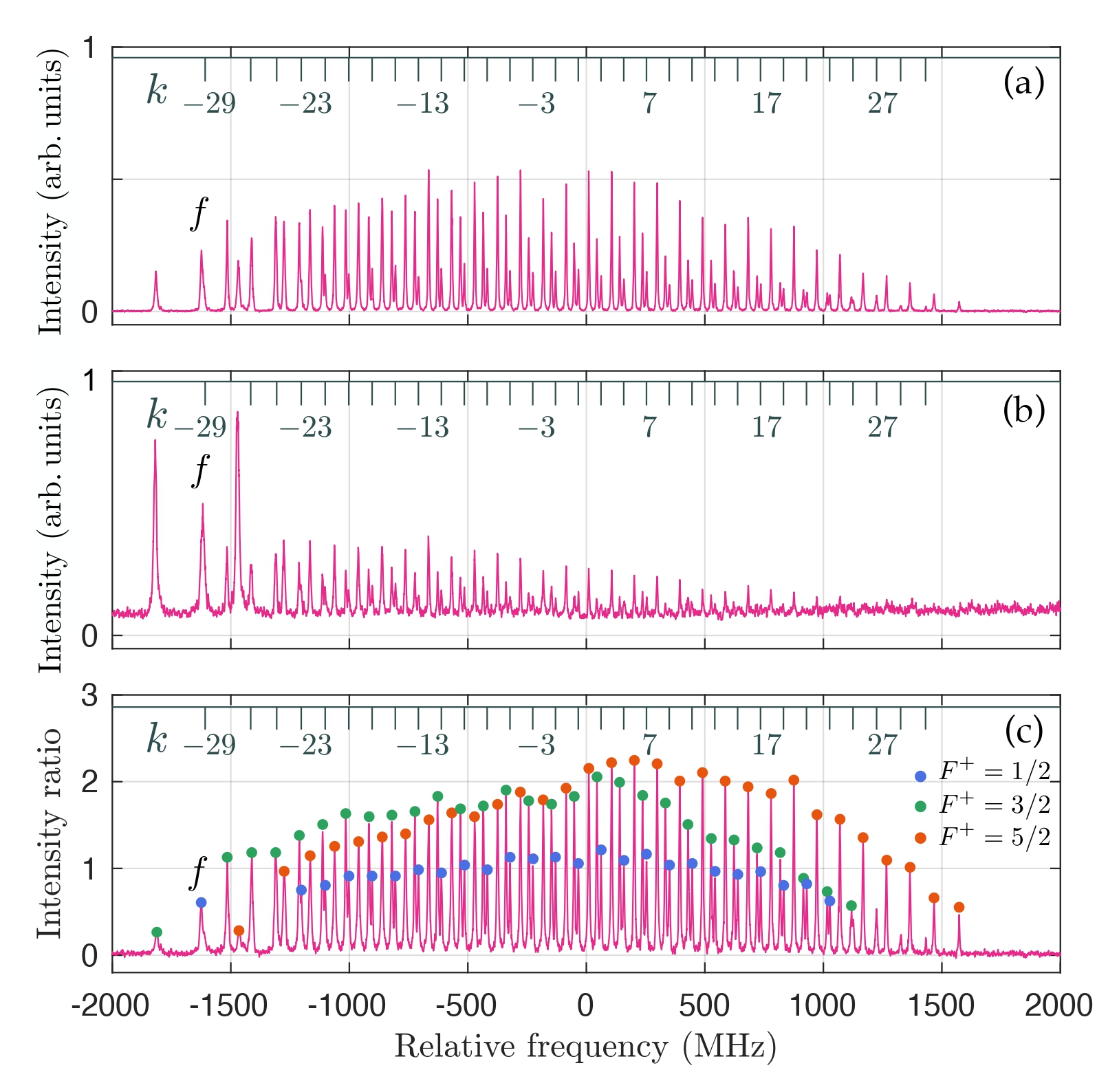}}
	\caption{Spectra of the $n=34$ [D$_2^+$ X$^+$(1,0)] Rydberg-Stark manifolds at electric field strengths of 69.79 V/m measured in the (a) slow and (b) fast-autoionization time-of-flight windows. (c): Ratio of the spectra in panels (a) and (b). The assignment bars give the value of $k$.}
	\vspace{-0.5cm}
	\label{fig3}
\end{figure}

The autoionization of the high-$k$ Rydberg-Stark states of D$_2^+$, which have the smallest contributions from short-lived low-$\ell$ states, was slow enough to lead to a tail in the D$_2^+$ time-of-flight distributions. Fig.~\ref{fig3} compares spectra obtained by recording the D$_2^+$ ion signal corresponding to the the tail (a) and the main signal (b) and their ratio (c). The comparison reveals that the $I=0$ Rydberg-Stark states (blue dots in Fig.~\ref{fig3}c) autoionize faster than the $I=2$ Rydberg-Stark states (green and orange dots for $F^+=3/2$ and 5/2, respectively). This dependence  of the autoionization on $I$ is not accounted for by the calculations because the MQDT parameters used in the model are the same for $I=0$ and 2.

Given the otherwise very good agreement between calculated and measured spectra (see Fig.~\ref{fig:spectra_calc_exp}), the calculations can be used to explain the different behaviors observed in H$_2$ and D$_2$. The calculated zero-field level structures at $n=30$ are depicted schematically in Fig.~\ref{fig1}a(iii), b(iii) and c(iii). The different $\ell$ levels are almost degenerate at zero field in the H atom \cite{tiesinga21a,scheidegger23a} (thick black bar in a(iii)). The Stark effect is thus almost perfectly linear. In D$_2$, the zero-field levels extend over a broader spectral region (gray and colored horizontal lines in b(iii)) because of the deviations from a pure Coulomb potential resulting from the charge distribution and the isotropic polarizability of the $N^+=0$ ion core. These deviations rapidly decrease with increasing $\ell$ values.  The Stark effect is similar to that of the H atom, with the main difference that the low-$\ell$ levels are subject to a quadratic Stark shift at low fields. 
The calculations reveal a negligible coupling of the nuclear ($I=2$) and core-electron ($S^+=1/2$) spins with the Rydberg electron orbital and spin angular momenta at high $n$.  As a result, the Rydberg electron motion is almost perfectly separable from the core motion that is dominated by the hyperfine interaction.
Consequently, the Rydberg-level structure is a superposition of three separate and almost identical manifolds for the $I=0,F^+=1/2$ (gray in Fig.~\ref{fig1}b(iii)), $I=2;F^+=3/2$ (violet) and $I=2;F^+=5/2$ (pink) core states, as observed in Fig.~~\ref{fig:spectra_calc_exp}a. 

In H$_2$ ($N^+=2$), additional zero-field splittings arise from the anisotropic core polarizability and the interaction between the electron and the ion-core quadrupole moment, which do not average out for $N^+\ge 1$ rotational levels [compare zero-field structures in Fig.~\ref{fig:spectra_calc_exp}b(iii) and c(iii)]. In this case, the core rotational angular momentum $\vec{N}^+$ is coupled to both the ion-core electron spin $\vec{S}^+$ via the spin-rotational interaction and the Rydberg-electron orbital angular momentum $\vec{\ell}$ via the charge-quadrupole interaction (see Fig.~\ref{fig1}c, top). The core motion is thus no longer separable from the Rydberg-electron motion, which leads to a more complex dynamical behavior and a less regular structure of the Stark states. 
The distinguishing feature between the two categories is thus the zero vs. nonzero value of $N^+$ and the role of the ion-core quadrupole in the latter case.
In both cases, the calculations quantitatively reproduce the observed structures and can be used to derive the ion-core level structures in least-squares fits.

\begin{figure}
    \centering
     {\includegraphics[trim=15cm 9cm 12cm 2cm, clip=true, width=\linewidth]{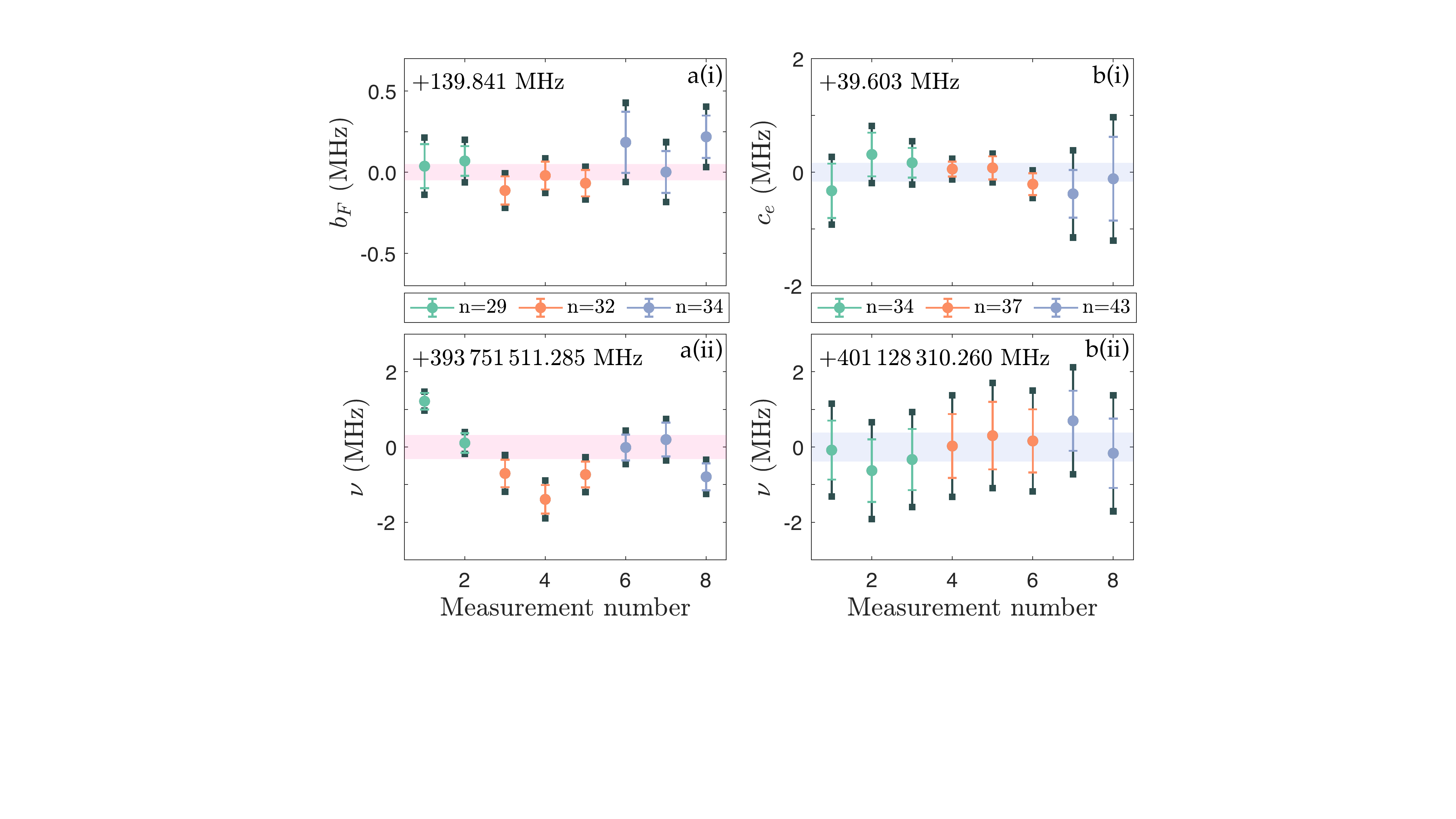}}
      \caption{Panels a(i) [b(i)]: Residuals of the fitted hyperfine (spin-rotation) coupling constants of D$_2^+$ (H$_2^+$). Panels a(ii) [b(ii)]: Residuals of the D$_2^+$ X$^+$(1,0) - GK(2,2) and H$_2^+$ X$^+$(1,2) - GK(2,0) intervals. The different data points at each value of $n$ indicate the results obtained from measurements at different values of the electric field strengths. The colored error bars show statistical uncertainties of the fitted parameters. Their black extensions represent systematic uncertainties of the calculations (see text for details). The colored areas indicate the final uncertainties.}
      \vspace{-0.5cm}
      \label{fig:resi}
\end{figure}
\begin{table}
\centering
\caption{Hyperfine (spin-rotation) coupling constants $b_F$ ($c_e$) in ortho-D$_2^+ (v^+=1, N^+=0)$ [para-H$_2^+ (v^+=1, N^+=2)$] and ionization energies D$_2^+$ X$^+$(1,0) - D$_2$ GK(2,2) [H$_2^+$ X$^+$(1,2) - H$_2$ GK(2,0)] and their uncertainties.}
\label{tab:rotint}
\begin{tabular}{ll} 
\toprule Energy interval & \ \ \ Value (MHz) \\
\midrule
$b_F$ [D$_2^+ (v^+=1,N^+=0)$] (Exp.$^\textrm{a}$) & \ \ \ 139.84(5)   \\
$b_F$ [D$_2^+ (v^+=1,N^+=0)$] (Th.$^\textrm{b}$) & \ \ \ 139.837(10) \\
D$_2^+$ X$^+$(1,0) - D$_2$ GK(2,2) (Exp.$^\textrm{a}$) & \ \ \  393\,751\,511.3(3) \\
\midrule
$c_e$ [H$_2^+$ X$^+ (1,2)$] (Exp.$^\textrm{a}$) & \ \ \ 39.62(11)   \\
$c_e$ [H$_2^+$ X$^+ (1,2)$]  (Th.$^\textrm{c}$)& \ \ \ 39.5716   \\
H$_2^+$ X$^+$(1,2) - H$_2$ GK(2,0) (Exp.$^\textrm{a}$) & \ \ \   401\,128\,310.3(3) \\
\bottomrule  
\end{tabular} 
\raggedright \\
{\small $^\textrm{a}$From this work. $^\textrm{b}$From Ref.\cite{danev21a}. $^\textrm{c}$ From Ref.\cite{korobov06c}.}
        \vspace{-0.5cm}
\label{tab:vals_final}
\end{table}
The ionization energies and the fine and hyperfine coupling constants obtained from these fits are presented in Table \ref{tab:vals_final} and the corresponding residuals for the measurements at different electric fields and $n$ values are displayed in Fig. \ref{fig:resi}. Typical error budgets for the cases of D$_2$ and H$_2$ are given in Tables SIII and SIV of the Supplemental Material \cite{supplemental_doran26b}. In each panel of Fig. \ref{fig:resi}, the different values of $n$ for which spectra were measured are indicated by different colors, and for each $n$ value the different data points correspond to different electric fields. The colored error bars represent statistical uncertainties (1$\sigma$). 
Their black extensions reflect systematic uncertainties in the quantum defects of the d series, which are the most strongly coupled to the f series and to the linear Stark manifolds. 
The shaded areas represent the final uncertainties encompassing both the contributions from the error bars of individual data points and their scatter \cite{meister00a}. The D$_2^+$ X$^+$(1,0) - D$_2$ GK(2,2) and H$_2^+$ X$^+$(1,2) - H$_2$ GK(2,0) ionization energies are determined at a relative accuracy better than 10$^{-9}$. The D$_2^+$ X$^+$(0,0) - D$_2$ GK(0,2) ionization energy from Ref. \cite{hussels22a} and the interval GK(2,2)-GK(0,2) of 42\,639\,437.7(9) MHz derived here in a separate measurement can be combined to determine a value of 47\,279\,980.8(1.9)~MHz for the fundamental vibrational frequency of D$_2^+$ [X$^+$(1,0) - X$^+$(0,0)], in perfect agreement with the more precise theoretical result of 47\,279\,981.589\,8(12)~MHz \cite{korobov17a}.

Precise experimental values for both the hyperfine coupling constant $b_F$ of D$_2^+$ ($v^+=1, N^+=0$) [139.84(5)~MHz] and the spin-rotation coupling constant $c_e$ of H$_2^+$ ($v^+=1, N^+=2$) [39.62(11) MHz] are reported here for the first time and agree with theoretical values within the experimental uncertainties, see Table \ref{tab:vals_final}. The accuracy of the present hyperfine levels of ortho-D$_2^+$ $(v^+=1, N^+=0)$ even surpasses the accuracy in the hyperfine structure of the nonautoionizing $v^+=0, N^+=1$ level of para-D$_2^+$ derived from millimeter-wave spectra of transitions between high-$n$ d and f Rydberg states under field-free conditions \cite{cruse08a}. 

The excellent agreement between experimental and calculated spectra achieved here validates the theoretical approach presented in Ref. \cite{iodo_tbp_first}. It enables the characterization of the structure and dynamics of high molecular Rydberg-Stark states with open-shell ion cores  at an unprecedented level and the determination of ionic fine- and hyperfine-structure intervals in molecular ions. The strength of this approach lies in its applicability to any molecular ion that is accessible by photoionization.
\\
\begin{acknowledgments}
We thank Dr. J. -Ph. Karr for useful advice on the theoretical treatment of the hyperfine structure of molecular hydrogen ions. We thank H. Schmutz and J. A. Agner for their technical assistance. This work is supported financially by the Swiss National Science Foundation (grant No.: 200021-236716) and the European Union’s Horizon Europe Research and Innovation Programme and the Participating States (funder ID: 10.13039/100019599, grant No.: 23FUN04 COMOMET).
\end{acknowledgments}
%

\begin{widetext}
\section*{Supplemental Material}
This supplemental material provides information on the fitted values of $\mathcal{F}_z$, the ionization energies D$_2^+$ X$^+ (1,0)$ - D$_2$ GK(2,2) [H$_2^+$ X$^+ (1,2)$ - H$_2$ GK(2,0)] and the hyperfine (spin-rotation) coupling constants $b_F$ ($c_e$) of ortho-D$_2^+ (v^+=1, N^+=0)$ [para-H$_2^+ (v^+=1, N^+=2)$] from the individual measurements of Rydberg-Stark manifolds at different $n$ values in Table \ref{tab:vals_fits_od2} (\ref{tab:vals_fits_ph2}). Typical error budgets accounting for systematic frequency corrections and uncertainties relevant for the determination of the ionization energies and hyperfine (spin-rotation) constants are given for D$_2$ and H$_2$ in Tables \ref{tab:errtr_d2} and \ref{tab:errtr_h2}, respectively.

\begin{table}[H]
\renewcommand{\thetable}{SI} 
\centering
\caption{Summary of fitted electric field strengths $\mathcal{F}_z$, ionization energy D$_2^+$ X$^+ (1,0)$ - D$_2$ GK(2,2) and hyperfine coupling constant $b_F$ in ortho-D$_2^+ (v^+=1, N^+=0)$ and their uncertainties (statistical and systematic, respectively, in the case of the ionization energy and of $b_F$), with the corresponding weighted RMS (=$\sqrt{\chi^2/\nu}$, where $\nu$ is the number of degrees of freedom) of the fits. Two consecutive fits are performed for each measurement at a given value of $n$ and $\mathcal{F}_z$: in the first fit, $\mathcal{F}_z$ and D$_2^+$ X$^+ (1,0)$ - D$_2$ GK(2,2) are adjusted to best match the transition frequencies to $I=0$ Rydberg-Stark states. These values of $\mathcal{F}_z$ and D$_2^+$ X$^+ (1,0)$ - D$_2$ GK(2,2) are kept constant in a second fit, in which $b_F$ is fitted using the transition frequencies to $I=2$ Rydberg-Stark states.}
\label{tab:rotint}
\begin{tabular}{lllll} 
\toprule  $n$ & \ \ \  $\mathcal{F}_z$ (mV/cm)  & \ \ \  D$_2^+$ X$^+ (1,0)$ - D$_2$ GK(2,2) (MHz)& \ \ \ $b_F$ (MHz) & \ \ \  RMS \\
\midrule
29 & \ \ \  1841.130(205)  & \ \ \  393\,751\,512.502(192)(30) &  & \ \ \  0.870 \\
29 &  & & \ \ \  139.880(93)(40) & \ \ \ 2.411  \\

29 & \ \ \ 1294.260(370) & \ \ \ 393\,751\,511.393(230)(30) & & \ \ \ 2.458 \\
29 & && \ \ \ 139.910(52)(40) & \ \ \ 3.363 \\

32 & \ \ \ 1195.100(248) & \ \ \ 393\,751\,510.583(245)(120) & & \ \ \ 0.563 \\
32 & & & \ \ \ 139.728(67)(20) & \ \ \ 2.162 \\

32 & \ \ \ 698.222(284) & \ \ \ 393\,751\,509.894(261)(120) & & \ \ \ 0.487 \\
32 & & & \ \ \ 139.820(66)(20) & \ \ \ 1.417\\

32 & \ \ \ 946.009(493) & \ \ \  393\,751\,510.553(224)(120) & & \ \ \ 1.369 \\
32 & & & \ \ \ 139.773(61)(20) & \ \ \ 2.635 \\

34 & \ \ \ 697.926(270) & \ \ \ 393\,751\,511.273(245)(100) &  &  \ \ \  2.663 \\
34 &  &  & \ \ \ 140.025(134)(55) &  \ \ \  4.570\\

34 & \ \ \ 797.271(331) & \ \ \ 393\,751\,511.482(350)(100) &  & \ \ \ 0.836 \\
34 & \ && \ \ \ 139.842(74)(55) & \ \ \ 2.700 \\

34 & \ \ \ 897.405(315) & \ \ \ 393\,751\,510.493(256)(100) & & \ \ \ 1.487 \\
34 &  & & \ \ \ 140.059(76)(55) & \ \ \ 3.748 \\

\bottomrule  
\end{tabular} 
\label{tab:vals_fits_od2}
\end{table}

\begin{table}[H]
\renewcommand{\thetable}{SII} 
\centering
\caption{Summary of fitted $\mathcal{F}_z$, ionization energy H$_2^+$ X$^+ (1,2)$ - H$_2$ GK(2,0) and hyperfine coupling constant $c_e$ in para-H$_2^+ (v^+=1, N^+=2)$ and their uncertainties (statistical and systematic, respectively, in the case of the ionization energy and of $c_e$), with the corresponding weighted RMS values of the fits.}
\label{tab:rotint}
\begin{tabular}{lllll} 
\toprule
$n$ & \ \ \  $\mathcal{F}_z$ (mV/cm)  & \ \ \  H$_2^+$ X$^+ (1,2)$ - H$_2$ GK(2,0) (MHz)& \ \ \ $c_e$ (MHz) & \ \ \  RMS \\
\midrule
34 & \ \ \ 1144.141(254)  & \ \ \  401\,128\,310.161(298)(450) & \ \ \ 39.314(169)(120) & \ \ \ 1.052 \\
34 & \ \ \ 1243.473(333) & \ \ \ 401\,128\,309.681(384)(450) & \ \ \ 39.927(264)(120) & \ \ \ 1.991 \\
34 & \ \ \ 1491.780(288) & \ \ \ 401\,128\,309.958(312)(450) & \ \ \ 39.772(288)(120) & \ \ \ 1.675 \\
37 & \ \ \ 1095.123(138) & \ \ \ 401\,128\,310.305(349)(500)  & \ \ \ 39.659(80)(50) & \ \ \ 1.322 \\
37 & \ \ \ 1194.058(269) & \ \ \  401\,128\,310.580(397)(500) & \ \ \ 39.680(153)(50) & \ \ \ 1.486\\
37 & \ \ \ 945.700(123) & \ \ \ 401\,128\,310.444(339)(500) & \ \ \ 39.400(142)(50) & \ \ \ 1.376 \\
43 & \ \ \ 995.269(102) & \ \ \ 401\,128\,311.262(176)(620) & \ \ \ 40.313(140)(350) & \ \ \ 2.038 \\
43 & \ \ \ 897.654(214) & \ \ \ 401\,128\,310.103(300)(620) & \ \ \ 39.491(386)(350) & \ \ \ 4.095 \\
\bottomrule  
\end{tabular} 
\label{tab:vals_fits_ph2}
\end{table}

\clearpage

\begin{table}
\renewcommand{\thetable}{SIII} 
\centering
\caption{Error budget and frequency corrections for the determination of the ionization energy D$_2^+$ X$^+$(1,0) - D$_2$ GK(2,2) and the hyperfine constant $b_F$ of ortho-D$_2^+ (v^+=1, N^+=0)$ from the measurement of transitions to the $n=34$ Rydberg-Stark manifold at $\mathcal{F}_z = 697.9$ mV/cm. All values and uncertainties are reported in kHz.}
\label{tab:errtr_d2}
\begin{tabular}{lll} 
\toprule
 & $\Delta \nu$  &  \ \ $\sigma $  \\
\midrule
Lineshape fits and  & & \\ 
residual 1$^ {\textrm{st}}$-order Doppler shift & & \ \ 250$^\textrm{a}$ \\ 
2$^ {\textrm{nd}}$-order Doppler shift & \ \ +2$^\textrm{b}$ &  \ \ 0.5 \\
ac-Stark shift &  & \ \ $<$5 \\
Zeeman shift &  &  \ \ $<$10 \\
Pressure shift &  &  \ \ $<$1 \\
Photon-recoil shift & $-$84$^\textrm{c}$  & \\
\bottomrule
\end{tabular}
\raggedright \\
{\small $^\textrm{a}$Only relevant for the determination of the ionization energy D$_2^+$ X$^+$(1,0) - D$_2$ GK(2,2) and averages out upon multiple realignments. $^\textrm{b}$Corresponds to a velocity of the D$_2$ beam of 950 m/s (valve temperature 80K). $^\textrm{c}$Corresponds to $\tilde\nu_{\textrm{laser}}$ =13\,039.221 cm$^{-1}$.}
\end{table}
\begin{table}
\renewcommand{\thetable}{SIV} 
\centering
\caption{Error budget and frequency corrections for the determination of the ionization energy H$_2^+$ X$^+$(1,2) - H$_2$ GK(2,0) and the hyperfine constant $c_e$ of para-H$_2^+ (v^+=1, N^+=2)$ from the measurement of transitions to the $n=34$ Rydberg-Stark manifold at $\mathcal{F}_z = 1243.5$ mV/cm. All values and uncertainties are reported in kHz.}
\label{tab:errtr_h2}
\begin{tabular}{lll} 
\toprule
 & $\Delta \nu$  &  \ \ $\sigma $  \\
\midrule
Lineshape fits and  & & \\ 
residual 1$^ {\textrm{st}}$-order Doppler shift & & \ \ 250$^\textrm{a}$ \\ 
2$^ {\textrm{nd}}$-order Doppler shift & \ \ +3$^\textrm{b}$ &  \ \ 0.5 \\
ac-Stark shift &  & \ \ $<$5 \\
Zeeman shift &  &  \ \ $<$10 \\
Pressure shift &  &  \ \ $<$1 \\
Photon-recoil shift & $-$175$^\textrm{c}$  & \\
\bottomrule
\end{tabular}
\raggedright \\
{\small $^\textrm{a}$Only relevant for the determination of the ionization energy H$_2^+$ X$^+$(1,2) - H$_2$ GK(2,0) and averages out upon multiple realignments. $^\textrm{b}$Corresponds to a velocity of the H$_2$ beam of 1200 m/s (valve temperature 60K). $^\textrm{c}$Corresponds to $\tilde\nu_{\textrm{laser}}$ =13\,285.298 cm$^{-1}$.}
\end{table}

\end{widetext}

\end{document}